\documentclass[12pt,preprint]{aastex}

\slugcomment{to appear in Astrophys. J.}

\shorttitle{Magnetically Supported Filamentary Clouds}
\shortauthors{Hanawa and Tomisaka}

\begin{document}


\title{Structure and Stability of Filamentary Clouds Supported  by Lateral Magnetic Field}


\author{Tomoyuki Hanawa}
\affil{Center for Frontier Science, Chiba University, Chiba, 263-8522, Japan}\email{hanawa@faculty.chiba-u.jp}

\and 
\author{Kohji Tomisaka\altaffilmark{1}}
\affil{Division of Theoretical Astronomy, National Astronomical Observatory of Japan, Mitaka, Tokyo 181-8588, Japan}
\email{tomisaka@th.nao.ac.jp}


\altaffiltext{1}{Also at the Department of Astronomical Science, School of Physical Sciences, SOKENDAI 
(Graduate University for Advanced Studies), Mitaka, Tokyo 181-8588, Japan}


\begin{abstract}
We have constructed two types of analytical models for an isothermal filamentary cloud 
supported mainly by magnetic tension.  The first one describes an isolated cloud while
the second considers filamentary clouds spaced periodically.  Both the models assume that
the filamentary clouds are highly flattened.   The former is proved to be the asymptotic limit 
of the latter in which each filamentary cloud is much thinner than the distance to the 
neighboring filaments.  We show that these models reproduce main features of the 2D
equilibrium model of \citet{tomisaka14} for filamentary cloud threaded by perpendicular magnetic field.
It is also shown that the critical mass to flux ratio is
$ M /\Phi  =  (2 \pi \sqrt{G}) ^{-1}  $, where $ M $, $ \Phi $ and $ G $ denote the cloud
mass, the total magnetic flux of the cloud, and the gravitational constant, respectively.  
This upper bound coincides with that for an axisymmetric cloud supported by
poloidal magnetic fields.  We applied the variational principle for studying
the Jeans instability  of  the first model.   Our model cloud is unstable against 
fragmentation as well as the filamentary clouds threaded by longitudinal
magnetic field.   The fastest growing mode has a wavelength several times 
longer than the cloud diameter.  The second model describes quasi-static evolution
of filamentary molecular cloud by ambipolar diffusion.
\end{abstract}





\section{Introduction}

Interstellar magnetic field plays significant roles on the dynamics of molecular clouds.
Magnetic field supports molecular cloud against gravitational collapse if it is strong enough.  
It is well known that the mass to flux ratio, i.e., the amount of gas contained in 
a magnetic flux tube, is the key parameter for the magnetic support.  \citet{mestel56} 
pointed out from the viral analysis that magnetic force reduces gravity by a constant factor
independent of the size of the cloud.   They evaluated the critical magnetic flux to support 
a cloud of mass, $ M $, 
to be $ \Phi _c =  \pi \sqrt{G} M $ \citep{mestel84}. 
The virial theorem provides the correct scaling law but not quantitatively correct value
for the critical mass.  One needs equilibrium models to evaluate the critical value.
\citet{mouschovias76a,mouschovias76b} for the first time obtained a magnetohydrostatic 
equilibrium for a disk like cloud supported in part by magnetic field.  He evaluated
magnetic pressure and tension as well as self-gravity and gas pressure consistently.
He confirmed that the mass to flux ratio is one of the key parameters.   He also noticed
that the equilibrium depends also on ratio of the magnetic pressure to the gas pressure 
and the flux function, i.e.,  the amount of gas contained in each magnetic flux.
\citet{tomisaka88} obtained a fitting formula for the critical equilibrium from their numerical
results.  It reads $ \Phi _c = (0.17) ^{-1} M \sqrt{G} \simeq 2 \pi \sqrt{G} M $.  
 Finite electrical conductivity changes the flux function through ambipolar diffusion 
and  induces quasi-static contraction.   When the magnetic field is weaker than the critical
one, the gravitational collapse continues to form stars.  Thus 
magnetohydrostatic equilibria help our understanding of the gravitational collapse
to form stars. 

Magnetohydrostatic equilibria depend on the configuration.  We need to consider 
filamentary clouds in magnetohydrostatic equilibria since observed molecular clouds
are often filamentary and associated with either longitudinal or perpendicular
magnetic field.  The magnetic field is parallel to elongation of the molecular cloud
in the Ophiuchus star forming region while it is perpendicular in the Taurus 
\cite[see, e.g.,][]{moneti84,goodman90,palmeirim14} and in the Musca \citep{pereyra04}.

\citet{stodolkiewcz63} and \citet{ostriker64} obtained equilibrium model having the
density distribution,
\begin{eqnarray}
\rho (r) & = & \rho _c \left( 1  + \frac{r ^2}{8H^2} \right) ^{-2} , \\
H ^2 & = & \frac{c _s ^2}{4 \pi G \rho _c} ,
\end{eqnarray}
for an isothermal filamentary cloud in the cylindrical coordinate, ($r, \varphi, z$),
where $ c _s $ and $ G $ denote the isothermal sound speed and gravitational
constant, respectively.  Interestingly, the line density defined by
\begin{eqnarray}
\lambda & \equiv & \int _0 ^{\infty} 2 \pi r  \rho (r) dr = \frac{2 c _s ^2}{G} ,
\end{eqnarray}
does not depend on the filament width, $ H $.  Thus, this model is neutrally
stable against radial contraction.   

\citet{stodolkiewcz63} showed that filamentary clouds having a larger line density 
can be sustained by purely longitudinal or partially helical magnetic field.
These models are unstable against fragmentation, i.e., sinusoidal perturbation
in the $ z $-direction \cite[see, e.g.,][]{hanawa93}.   The fragmentation and its
further evolution were studied extensively by \citet{tomisaka95}, \citet{nakamura95} and
others.

On the contrary, filamentary clouds permeated by magnetic field perpendicular to 
the axis have been studied little.  Its structure was obtained only recently by
\citet{tomisaka14} referred to Paper I in the following.  This is mainly because
the structure is obtained only by extensive numerical computation solving the Poisson 
and Grad-Shafranov equations simultaneously.  The latter is highly nonlinear 
since it describes force balance in the directions parallel and perpendicular
to the magnetic field.   It requires high spatial resolution especially when
the filamentary cloud is condensed.   Paper I showed that the maximum line
density increases in proportion to the magnetic flux.  In this paper we show 
analytical models  which reproduce  the main features of the 2D models 
obtained in Paper I.  

When the line density is higher than $ 2 c _s ^2 / G $ and thus supported 
mainly by magnetic field, the filamentary cloud is flattened like Italian pasta 
fettucine.  The magnetic force is dominated by the magnetic tension.  Then 
we can use the thin disk approximation to analyze the structure and stability
of filamentary cloud permeated by perpendicular magnetic field.

This paper is organized as follows.  In \S 2 we derive an analytical model for
an isolated filamentary cloud.   The mass to flux ratio is uniform in the model.
In \S 3 we use the Fourier series for filamentary clouds arranged periodically.
It is shown that the solution approaches to that obtained in \S 2 when the
filament thickness is much smaller than the distance to the neighboring 
filaments.   It is also shown that the ratio of the thickness to the distance
can be regarded as the model parameter specifying the quasi-static evolution.
In \S 4 we confirm that our 1D models reproduce the main features
of the 2D model obtained in Paper I.  In \S 5 the model obtained in \S 2 is
proved to be unstable against fragmentation.  In \S 6 we discuss the stability of the
models obtained in \S 3.

\section{Equilibrium Model}

In this paper, we consider an isothermal flattened filamentary cloud having infinite length and 
infinitesimal thickness.  This simplification is justified for observed filamentary clouds
since their lengths are much larger than the widths and the magnetic fields are
often perpendicular to cloud elongation on the sky.   It is well known that self-gravitating
clouds are flattened in the direction perpendicular to the global magnetic field.
In the following we use the Cartesian coordinates in which the cloud is confined in
the plane of $ z = 0 $ and elongated in the $ y $-direction.  Both the gas and 
magnetic fields are uniform in the $ y $-direction.   

From the above assumptions we can express the gas density distribution as
\begin{eqnarray}
\rho & = & \Sigma (x) \, \delta (z) ,
\end{eqnarray}
where $\Sigma (x) $ and  $ \delta (z) $ denote the surface density and the Dirac's 
delta function, respectively.   The Poisson equation reduces to
\begin{eqnarray}
\left( \frac{\partial ^2}{\partial x^2} + \frac{\partial ^2}{\partial z ^2} \right)
\phi & = & 4 \pi G \Sigma (x) \delta (z) . \label{poisson}
\end{eqnarray}
We obtain the boundary condition,
\begin{eqnarray}
\left. \frac{\partial \phi}{\partial z} \right| _{z=+\varepsilon} & = & - 
\left. \frac{\partial \phi}{\partial z} \right| _{z=-\varepsilon} = 2 \pi G \Sigma (x) . 
\label{g-boundary}
\end{eqnarray}
by integrating Equation (\ref{poisson}) in the vertical direction over the
infinitesimal interval around $ z = 0 $
\cite[see, e.g.,][for the derivation]{binney08}.  The symbol,
$ \varepsilon $, denotes an infinitesimally small positive quantity.  
The gravity is given by
\begin{eqnarray}
g _x & = & - \frac{\partial \phi}{\partial x} , \\
g _z & = & - \frac{\partial \phi}{\partial z} .
\end{eqnarray}

We assume for simplicity that the magnetic field has only the $ x $- and 
$ z $-components in the equilibrium.    
The condition for magneto-hydrostatic equilibrium requires that  the magnetic field 
should be force free and hence current free outside the disk,
\begin{eqnarray}
\frac{\partial B _x}{\partial z} - \frac{\partial B _z}{\partial x} & = & 
\frac{4 \pi J _y (x)}{c} \delta (z) ,
\end{eqnarray}
since both the  gravity and pressure thereof are negligibly small.  
The symbols, $ c $ and $ J _y (x) $, denote the speed of light and  the electric surface 
current running in the  $ y $-direction inside the cloud per unit length in the $ x $-direction,
respectively.   The magnetic field is kinked on the disk,
\begin{eqnarray}
B _x (x, z=+\varepsilon) & = & - B _x (x, z=-\varepsilon) = \frac{2 \pi J _y (x)}{c} . 
\end{eqnarray}

Note that the gravity and  magnetic field can be perfectly aligned outside the cloud.
Both of them are divergence and rotation free, i.e., 
$ \mbox{\boldmath$\nabla$} \cdot \mbox{\boldmath$g$} = 0 $,
$ \mbox{\boldmath$\nabla$} \times \mbox{\boldmath$g$} = 0 $,
$ \mbox{\boldmath$\nabla$} \cdot \mbox{\boldmath$B$} = 0 $, and
$ \mbox{\boldmath$\nabla$} \times \mbox{\boldmath$B$} = 0 $,
outside the cloud,  $ z \ne 0 $.  In the following
we focus on the case that the magnetic field is proportional to the gravity,
\begin{eqnarray}
B _x & = & - \frac{\alpha}{\sqrt{G}} g _x , \\
B _z & = & - \frac{\alpha}{\sqrt{G}} g _z , \label{isopedic}
\end{eqnarray}
in $ z > 0 $ and 
\begin{eqnarray}
B _x & = & \frac{\alpha}{\sqrt{G}} g _x , \\
B _z & = & \frac{\alpha}{\sqrt{G}} g _z , \label{isopedic2}
\end{eqnarray}
in $ z < 0 $,  where $ \alpha $ denotes a non-dimensional constant.   This assumption is equivalent
to  the constant mass to flux ratio \citep[isopedic in the terminology of][]{shu97},
\begin{eqnarray}
\alpha  & = & \frac{B _z (x, z=0)}{2 \pi \sqrt{G} \Sigma (x)} , \label{masstof}
\end{eqnarray}
thanks to Equation (\ref{g-boundary}).   
Hence, the cloud is subcritical (supercritical)
when $ \alpha > 1 $ ($\alpha < 1 $).
When the mass to flux ratio is constant, the magnetic tension working on the cloud is
proportional to the gravity,
\begin{eqnarray}
\int _{-\varepsilon} ^{+\varepsilon}  
\frac{\partial}{\partial z} \left( \frac{ B _x  B_z}{4\pi} \right) dz & = & 
\left. \frac{B _x B _z }{2 \pi} \right| _{z=+\varepsilon} \\
& = & - \alpha ^2 g _x \Sigma , \label{magf}
\end{eqnarray}
in the cloud.   

Equations (\ref{magf}) simplifies the equation for magnetohydrostatic balance in
the $ x $-direction,
\begin{eqnarray}
\frac{\partial}{\partial x} \left(  c _s ^2 \rho \right) + \frac{\partial}{\partial x}
\left( \frac{B _x ^2 + B _z ^2}{8 \pi} \right) - \frac{\partial}{\partial z} \left( 
\frac{ B _x B _z }{4 \pi} \right)  & = & g _x \rho ,  \label{MHDvertical} 
\end{eqnarray}
where $ c _s $ denotes the isothermal sound speed.
Substituting Equation (\ref{magf}) into Equation (\ref{MHDvertical}) integrated over the 
infinitesimal interval around $ z = 0 $, we obtain
\begin{eqnarray}
\left(1 - \alpha ^2 \right) g _x \Sigma + c _s ^2 \frac{\partial \Sigma}{\partial x} & = & 0 .
\label{equilibrium}
\end{eqnarray}
The magnetic field reduces the effective gravity by a factor $ \left(1 - \alpha ^2 \right) $.
See \citet{shu97} for the validity of this approximation.
Equation (\ref{equilibrium}) indicates that the cloud is in equilibrium only when 
$ \alpha < 1 $ (supercritical).

In the following we consider the Lorentz profile,
\begin{eqnarray}
\Sigma = \frac{\lambda}{\pi} \frac{a}{x ^2 + a ^2} \label{surface}
\end{eqnarray}
as a model for the flattened filamentary cloud.   The line density and FWHM of this model
cloud are  $ \lambda $ and $ 2 a $, respectively.   Using the method of images we 
obtain the corresponding gravity,
\begin{eqnarray}
\mbox{\boldmath$g$} = 
\left( \begin{array}{c} g _x \\ g _z \end{array} \right) & = & 
- \frac{z}{|z|} \frac{ 2 G \lambda}{x ^2 + (|z|+a) ^2}  
 \left( \begin{array}{c} x \\ 
\displaystyle |z| + a \end{array} \right) .  
\end{eqnarray} 
The pressure force is given by
\begin{eqnarray}
c  _s ^2 \frac{d \Sigma}{dx} & = & - \frac{2 a c _s ^2 \lambda x}{\pi \left[ x ^2 + (|z| + a) ^2 \right] ^2} .
\label{pressure}
\end{eqnarray}
We obtain the condition for the magnetohydrostatic equilibrium,
\begin{eqnarray}
G \lambda \left( 1 - \alpha ^2 \right) - c _s ^2 & = & 0 ,  \label{filament}
\end{eqnarray}
by substituting Equations (\ref{surface}) through (\ref{pressure}) to Equation (\ref{equilibrium}).
Equation (\ref{filament}) is rewritten as
\begin{eqnarray}
\lambda = \frac{c _s ^2}{2G} + \sqrt{ \displaystyle \left( \frac{\Phi}{2 \pi \sqrt{G}} \right) ^2
+ \left( \frac{c  _s ^2}{2 G} \right) ^2 } ,  \label{lined}
\end{eqnarray}
by the help of Equation (\ref{masstof}).   
When Equation (\ref{lined}) holds, the model
cloud settles in an equilibrium for any $ a $.   In other words, the width of the cloud is
not specified by $ \lambda $ or $ \Phi $.  This character is common for isothermal
filamentary cloud in equilibrium.

The magnetic field is expressed as
\begin{eqnarray}
\mbox{\boldmath$B$} = 
\left( \begin{array}{c} B _x \\ B _z \end{array} \right) & = & 
\frac{ \Phi }{ \pi [ x ^2 + (|z|+a) ^2]} \left( \begin{array}{c} \displaystyle \frac{zx}{|z|} \\ 
\displaystyle |z| + a \end{array} \right) .  
\end{eqnarray} 
The magnetic field lines are straight and kinked on the plane of the flattened cloud.
They look as if they emanate radially from the lines of $ (x,z)=(0,\pm a)$.
The strength decreases inversely proportional to the distance from the lines.

We can estimate the density by assuming the hydrostatic balance in the 
vertical direction,
\begin{eqnarray}
\rho (x,z) & = & \rho (x,0) \exp \left\{ c _s ^{-2} \left[ \phi (x,0) - \phi (x,z) \right] \right\} , \\
\phi & = & G \lambda \ln \left( x ^2 + z ^2 + a ^2 \right) .
\end{eqnarray}
Here the gravitational potential, $ \phi $, is evaluated by the method of image and the
magnetic force is neglected in the hydrostatic balance.   The magnetic force,
$ \mbox{\boldmath$j$} \times \mbox{\boldmath$B$} $, vanishes outside the disk
since the electric current is confined in the disk.   The magnetic force is perpendicular
to the field inside the disk.  Hence it compresses the gas disk 
\citep[see][for the evaluation of the magnetic force acting on  the vertical structure]{shu97}.   
The condition,
\begin{eqnarray}
\Sigma (x) & = & \int _{-\infty} ^{+\infty} \rho (x, z) dz ,
\end{eqnarray}
gives us
\begin{eqnarray}
\rho (x,0) & = & \frac{\Sigma}{\sqrt{x^2+a^2}} \frac{\Gamma (G \lambda c _s ^{-2})}
{\sqrt{\pi} \Gamma \left(G \lambda c _s ^{-2} - 1/2\right) } ,
\end{eqnarray}
where $ \Gamma $ denotes the gamma function.  Accordingly we have
\begin{eqnarray}
\rho (x,z) & = & \frac{\lambda a}{\pi} \frac{\Gamma (G \lambda c _s ^{-2})}
{\sqrt{\pi} \Gamma \left(G \lambda c _s ^{-2} - 1/2\right) }
 \frac{\left( x ^2 + a ^2 \right) ^{G \lambda c _s ^{-2} -3/2}}
{\left( x ^2 + z ^2 + a^2 \right) ^{G \lambda c _s ^{-2}}}
\end{eqnarray}

\begin{figure}
\plottwo{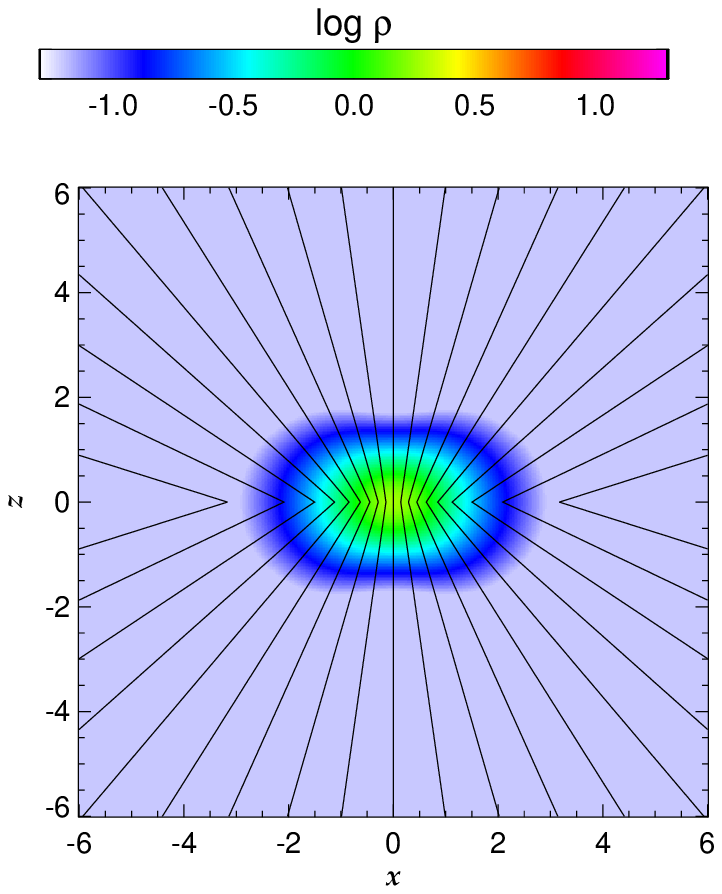}{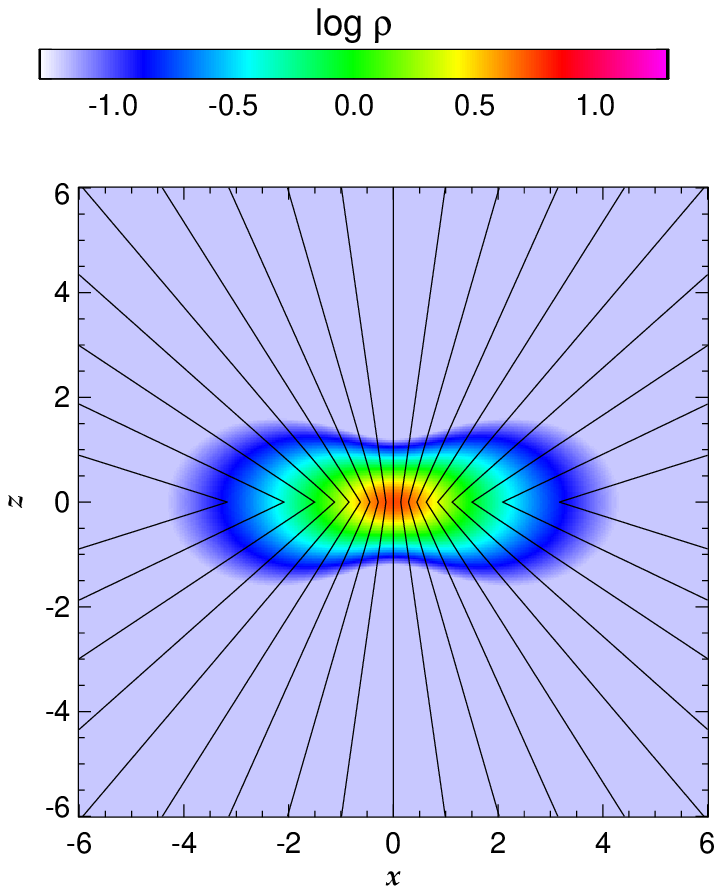}
\epsscale{.50}
\plotone{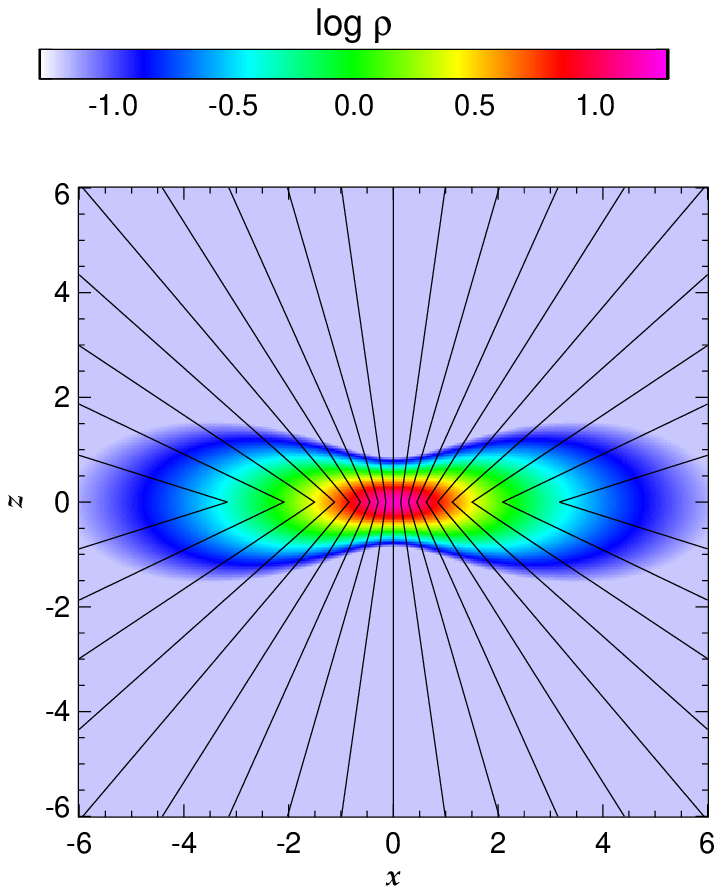}
\caption{The cross sections show the density by color and the magnetic
field by lines.  The upper left and right panels are for $ \lambda = 5 $ and  $ 10 c _s ^2 / G $,
respectively.  The bottom one is for $ \lambda = 20 c _s ^2 / G $. \label{rho-xz}}
\end{figure}

Figure \ref{rho-xz} denotes the density distribution and the magnetic field for
$ \lambda $ = 5, 10,  and 20 $ c _s ^2 / G $.   Here the abscissa and ordinate
are denoted in unit of $ a $ while the density is shown in unit of
$ c _s ^2 /(Ga) $.   The cloud is more flattened  for a larger $ \lambda $.

\section{Periodically Arranged Filamentary Clouds}

In this section we relax the uniform mass to flux ratio, $ \Sigma/ B  = \mbox{const.}$, 
assumed in the previous section.  The mass to flux ratio should  increase at the cloud
center through the ambipolar diffusion if the cloud is supported in part by the magnetic field
against gravity.   On the contrary, the mass to flux ratio decreases outside the cloud.
In order to take account of magnetic field outside clouds, we consider the situation 
in which filamentary clouds are arranged periodically on a plane.  Again each
filamentary cloud is assumed to be geometrically thin.  Then the surface density  is given
by the Fourier series,
\begin{eqnarray}
\Sigma (x) & = & \frac{1}{\ell} \sum _{n=0} ^\infty a _n \cos \left( \frac{2 \pi n x}{\ell} \right),
\label{psigma}
\end{eqnarray}
where $ \ell $ denotes the spacing between the filaments.   The 0th Fourier series coefficient,
$ a  _0 $, denotes line density per each filament, $ \lambda $.
We obtain the gravitational
potential,
\begin{eqnarray}
\phi (x,z) & = & \frac{2 \pi G a _0}{\ell} \vert z \vert - \sum _{n=1} ^\infty 
\frac{G a _n}{n}  \cos \left( \frac{2 \pi n x}{\ell} \right) \exp
\left( - \frac{2 \pi n}{\ell} |z| \right)
\end{eqnarray}
by solving the Poisson equation.  The gravity is expressed as
\begin{eqnarray}
g _x & = & - \frac{2 \pi G}{\ell} \sum _{n=1} ^\infty a _n \sin \left( \frac{2\pi n x}{\ell} \right) 
\exp \left( - \frac{2 \pi n}{\ell} |z| \right)  , \\
g _z & = & - \frac{2 \pi G}{\ell} \frac{z}{|z|} \sum _{n=0} ^\infty a _n \cos \left( \frac{2\pi n x}{\ell} \right) 
\exp \left( - \frac{2 \pi n}{\ell} |z| \right)  .
\end{eqnarray}
Similarly the force free magnetic field is expressed as
\begin{eqnarray}
B _x & = & - \frac{1}{\ell} \frac{z}{|z|} \sum _{n=1} ^\infty  b _n \sin \left( \frac{2\pi n x}{\ell} \right) 
\exp \left( - \frac{2 \pi n}{\ell} |z| \right)  , \\
B _z & = & \frac{1}{\ell} \sum _{n=0} ^\infty  b _n \cos \left( \frac{2\pi n x}{\ell} \right) 
\exp \left( - \frac{2 \pi n}{\ell} |z| \right)  .  
\end{eqnarray}
The 0th Fourier series coefficient, $ b _0 $, denotes the magnetic flux permeating each
filament, $ \Phi $ in the model for an isolated filamentary cloud.  

The condition for the force balance,
\begin{eqnarray}
c  _s ^2 \frac{\partial \Sigma}{\partial x} -  \left. \frac{B _x B _z}{2 \pi} \right| _{z=+\varepsilon} 
& = & g _x \Sigma ,
\label{balance}
\end{eqnarray}
is rewritten as
\begin{eqnarray}
c _s ^2 n a _n + \frac{1}{8 \pi ^2} \sum _{m=0} ^n 
b _m b _{n-m}  = \frac{G}{2} \sum _{m=0} ^n a _m a _{n-m} . \label{balance2}
\end{eqnarray}
This equation can be solved successively if once $ \mbox{\boldmath$a$} \equiv (a _1, a _2,
\dots) $ and $ b _0 $ are given.
The Fourier series coefficient, $ b _n $, is given by
\begin{eqnarray}
b _n & = & 
\frac{8 \pi G}{b _0}
\left( \frac{G}{2} \sum _{m=0} ^n
a _m a _{n-m} - c _s ^2 n a _n - \frac{1}{8 \pi G}
\sum _{m=1} ^{n-1} b _m b _{n-m} \right)  ,  \label{period-eq}
\end{eqnarray}
if $ a _0$, $ \dots,  a _n $, $ b _0 $, $\dots, b _{n-1} $ are known.  In other words we can
solve Equation (\ref{balance}) if the mass distribution and the magnetic flux are given.

In the following we consider the case,
\begin{eqnarray}
a _j  & = & \left\{ \begin{array}{ll} \lambda  & (j = 0) \\
2 \lambda \exp ( - w  j ) & (\mbox{otherwise}) 
\end{array} \right. , \label{seriesA}
\end{eqnarray}
where $ w $ denotes the ratio of the filament width to the interval.
 Then Equation (\ref{balance2}) reduces to
\begin{eqnarray}
\frac{1}{8 \pi ^2} \sum _{m=0} ^n b _m b _{n-m} & = & 2 n \lambda e ^{-n w} \left( 
G \lambda - c _s ^2 \right) .
\end{eqnarray}
The right hand side of this equation indicates that the gravity overcomes the pressure
force when $ \lambda > c _s ^2 / G $.

After some
algebra we find a solution,
\begin{eqnarray}
b _j  & = & \left\{ \begin{array}{ll} \Phi  & (j = 0) \\
2 \Phi \exp ( - w  j ) & (\mbox{otherwise}) 
\end{array} \right. , \\
\frac{\Phi ^2}{4 \pi ^2} & = & G \lambda ^2 -  c _s ^2  \lambda .
\end{eqnarray}
The mass to flux ratio is constant in this solution.
The surface density and magnetic fields are expressed as
\begin{eqnarray}
\Sigma (x,z=0) & = & \frac{\lambda \sinh w}
{\ell \left[ \cosh w - \cos \left( \displaystyle \frac{2 \pi x}{\ell} \right) \right]} , \label{period1} \\
B _x (x,z=+\varepsilon) & = & \frac{\Phi \sin \left( \displaystyle \frac{2\pi x}{\ell} \right) } 
{\ell \left[ \cosh w - \cos \left( \displaystyle \frac{2 \pi x}{\ell} \right) \right]} , \\
B _z (x,z=0) & = & \frac{\Phi \sinh w}
{\ell \left[ \cosh w - \cos \left( \displaystyle \frac{2 \pi x}{\ell} \right) \right]} , \label{period3}
\end{eqnarray}
respectively.  See Appendix A for more details on the derivation.
This solution of $ w = 2 \pi a /\ell  $ approaches to that given in \S 2 for 
a given $ a $  in  the limit of $ \ell \rightarrow \infty $ and hence $ w \rightarrow 0 $.

The solution obtained in the previous paragraph is a critical one.   
When $ b _0 $ is larger than the critical value,
\begin{eqnarray}
b _0 > 2 \pi \sqrt{G} \sqrt{ a _0 \left( a  _0 - \displaystyle \frac{c _s ^2}{G} \right)} . 
\label{criticalB}
\end{eqnarray}
we obtain solutions  in which the mass to flux ratio is not uniform.
Equation (\ref{criticalB}) can be rewritten as
\begin{eqnarray}
a _0 < \frac{c _s ^2}{2 G} + \sqrt{\left( \frac{b _0}{2 \pi \sqrt{G}} \right)  ^2
+ \left( \frac{c _s ^2}{2 G} \right) ^2} ,  \label{criticalB2}
\end{eqnarray}
the right hand side of which denotes the maximum line density supported by
magnetic flux, $ b _0 $.  The filamentary clouds are confined by the magnetic
pressure of the neighboring clouds.  Remember that the filamentary cloud
is confined by uniform magnetic field in the 2D model of Paper I.  Thus 
the model shown in this section is closer to the 2D model than that in the
previous section. 

\begin{figure}
\epsscale{.50}
\plotone{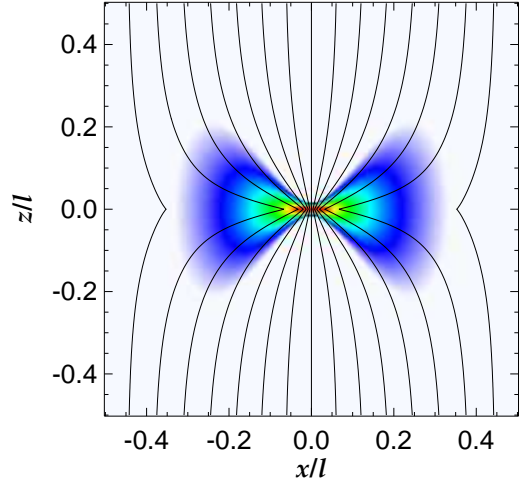}
\caption{The cross section shows the logarithm of the density by color and the magnetic
field by lines for $ \lambda = 5 c _s ^2 / G$ and  $ \Phi = 10 \pi  c _s ^2 / \sqrt{G} $.   
respectively.   \label{rho-xz2}}
\end{figure}

Figure \ref{rho-xz2} shows the magnetic field for $ \lambda = 5 c _s ^2 / G $ and
$ b _0 = 10 \pi c _s ^2 / \sqrt{G} $.   The color denotes the density evaluated to be
\begin{eqnarray}
\rho  (x, z) & \equiv & \frac{\pi G \Sigma (x) ^2}{2 c _s ^2} 
\left\{ \cosh \left[ \frac{\pi G \Sigma (x)  z}{c _s ^2} \right] \right\} ^{-2} , \label{rho}
\end{eqnarray}
in the logarithmic scale.  Equation (\ref{rho}) denotes the equilibrium density distribution
for an isothermal plane parallel disk having the surface density, $ \Sigma (x) $.  
The magnetic field is vertical to the disk plane near 
$ | x | = \ell / 2 $.

\begin{figure}
\epsscale{0.50}
\plotone{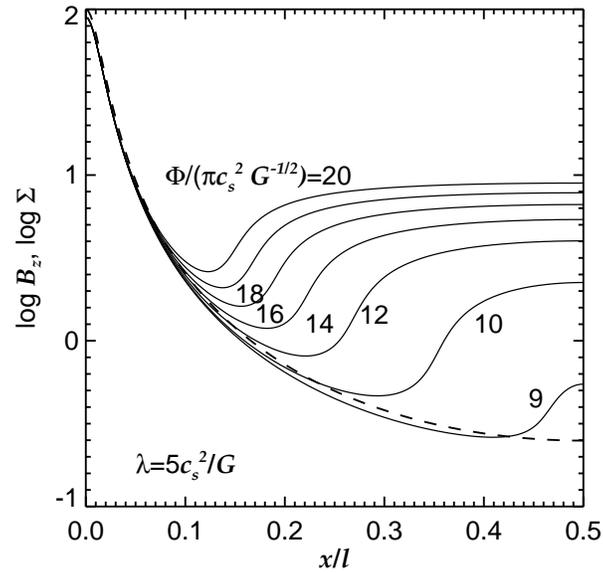}
\caption{The solid curves denote the distribution of $ B _z $ for $ \lambda = 5 c _s ^2 / G $.
The surface density is denoted by the dashed curve.
\label{lambda5}}
\end{figure}

Figure \ref{lambda5} shows the distribution of $ B _z $ for models having various 
magnetic fluxes, $ \Phi = b _0 = 9 $, 10, 12, 14, 16, 18, and 20 $\pi c _s ^2 /\sqrt{G} $. 
All the models have the same line density ($ \lambda = 5 c _s ^2 / G $)
and accordingly  the same surface density distribution shown by the dashed curve.
The value of $ w $ is fixed at $ w = 0.1 $.
The magnetic field and the surface density profiles are very similar near the
filament axis ($ x \simeq 0 $).  In other words almost the same amount of magnetic flux is
confined in the filament and the mass to flux ratio is constant, $ B _z = 2 \pi \sqrt{G} \Sigma $.
When the magnetic flux is larger, the magnetic field
is stronger outside the filamentary cloud.   When the magnetic flux is less than
a critical value ($ b _0 < 9 \pi c _s ^2 /\sqrt{G} $ in case of $ \lambda = 5 c _s ^2 / G $, 
see Eq. (\ref{criticalB})), 
we cannot construct an equilibrium model.   When it is close to the critical value,
the mass to flux ratio is nearly constant also outside the cloud as in the model
shown in \S 2. 

\begin{figure}
\epsscale{0.50}
\plotone{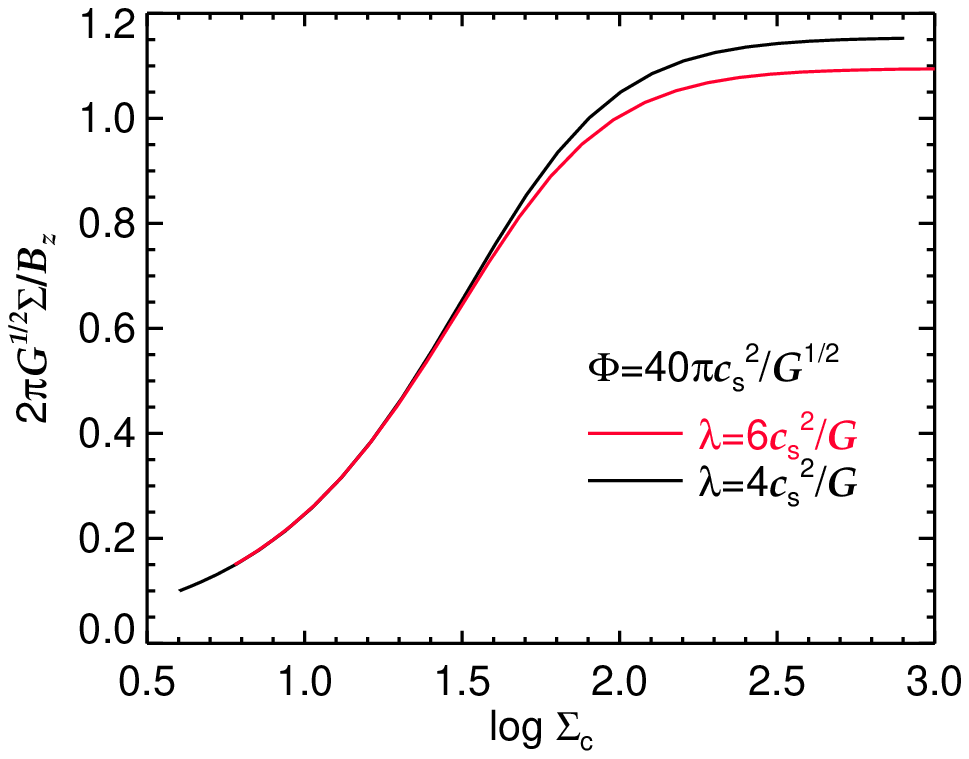}
\caption{The mass to flux ratio at the cloud center is shown as a function of the 
surface density at the cloud center for given $ \lambda $ and $ \Phi $,
which are taken to be $ \lambda = 4 c  _s ^2/G $ and $ \Phi = 8 \pi c _s ^2 / \sqrt{G} $
for the black curve and   $ \lambda = 6 c  _s ^2/G $ and $ \Phi = 8 \pi c _s ^2 / \sqrt{G} $
for the red curve, respectively.
\label{Sc_ratio}}
\end{figure}

Next we compare the models having the same $ \lambda $ and $ \Phi $ but different  $ w $.
Decrease in $ w $ mimics quasi-static contraction since the surface density at the cloud center
increases.   Figure \ref{Sc_ratio} denotes the mass to flux ratio at the cloud center, 
$  2 \pi \sqrt{G} \Sigma _c / B _{z,c} $, as a function of $ \Sigma _c $, where
$ \Sigma _c $ and $ B _{z,c} $ denote the surface density and magnetic field at
$ x = 0 $, respectively.   The ordinate is denoted in unit of $ c _s ^2 /(G \ell) $.
The red curve denotes the models having $ \lambda = 6 c _s ^2 / G $
and $ \Phi = 40 \pi c _s ^2 / \sqrt{G} $ while the black one does those $ \lambda = 4 c _s ^2 / G $
and $ \Phi = 40 \pi c _s ^2 / \sqrt{G} $.   As the surface density at the cloud center increases,
the mass to flux ratio there increases monotonically and is saturated at
\begin{eqnarray}
\left. \frac{2 \pi \sqrt{G} \Sigma}{B _z} \right| _{\Sigma \rightarrow \infty} 
& = & \left( 1 - \frac{c _s ^2}{G \lambda} \right) ^{-1/2} .
\end{eqnarray}
Note that the mass to flux ratio tends to $ \lambda / \Phi $ in the limit of $ w \rightarrow \infty $
since both the surface density and magnetic field are uniform in the limit.   
Interestingly, however, the mass to flux ratio at the cloud center depends little on $ \lambda $ 
for an intermediate $ w $.

When the line density is smaller than the critical value, $ c _s ^2 /  G $, the filamentary
cloud is confined not by the self-gravity but by the magnetic field.   This is because the
gas pressure dominates over the gravity.    We can construct
such a model by assuming a small $ \lambda $ and a large magnetic flux.    
They are similar to models with small $ R _0 $ in paper I. 

\section{Comparison with 2D model}

The thin disk model shown in \S 2 and 3 reproduces main features of the 2D model shown in Paper I.

Equation (\ref{filament}) is essentially the same as Equation (45) of Paper I,
\begin{eqnarray}
2 \pi R ^2 p _s = 2 c _s ^2 \lambda - G \lambda ^2  + \frac{\Phi _{\rm cl} ^2}{8\pi} , \label{tomisaka45}
\end{eqnarray}
where $ R $ and $ p _s $ denote the radius of the filament and the gas pressure at cloud surface,
respectively.
The symbol, $ \Phi _{\rm cl} $, denotes a half of the magnetic flux permeating the filament per unit
length.   Equation (\ref{tomisaka45}) was derived from the virial analysis and the left hand side denotes the
pressure force acting on the cloud surface, which vanishes in our 1D model.   Differences in the numerical
factors come from the assumptions used.   In this paper we neglected the finite thickness of the cloud
and hence pressure force acting in the $ z $-direction for simplicity.   Thus the critical line density is evaluated
to be $ \lambda _{cr} = c _s ^2 / G $ in case of no magnetic field ($ \Phi = 0 $) in this paper while
it should be $ 2 c _s ^2 / G $ in paper I.   Another difference comes from the assumed mass to flux ratio; it is
uniform in the model shown in \S 2 while it is not in Paper I.

The following equation,
\begin{eqnarray}
\lambda & = & \frac{c _s ^2 + \left( c _s ^4 + G \Phi _{\rm cl} ^2 / 8 \right) ^{1/2}}{G} , \label{tomisaka46}
\end{eqnarray}
was derived from Equation (45) of Paper I as well as Equation (\ref{lined}) is derived from
Equation (\ref{filament}).   Equation (\ref{tomisaka46}) has an asymptotic form,
\begin{eqnarray}
\lambda & = & \frac{c _s ^2}{G} + \frac{\Phi _{\rm cl}}{2 \sqrt{2 G}} + {\cal O} \left( \Phi _{\rm cl} ^{-1} \right) ,
\end{eqnarray}
which resembles Equation (38) of Paper I,
\begin{eqnarray}
\lambda _{\max} \simeq 0.24 \frac{\Phi _{\rm cl}}{\sqrt{G}} + 1.66 \frac{c _s ^2}{G} . 
\label{tomisaka38}
\end{eqnarray}
Equation (\ref{tomisaka38}) is useful since it provides an upper limit on the line density supported
against gravity  by magnetic field and gas pressure.  The upper limit increases in proportion to
the magnetic flux contained in the cloud.  Thus we examine the coefficient quantitatively.

In paper I the mass to flux ratio is assumed to be the same as that of the filament having
uniform density, $ \rho _0 $, and threaded by uniform magnetic field, $ B _0 $.  
Then the line density and magnetic flux are denoted by
\begin{eqnarray}
\lambda & = & \pi \rho _0 R _0 ^2 , \\
\Phi _{\rm cloud} & = & 2 B _0 R _0 ,  \label{Phi2D}
\end{eqnarray}
respectively, where  $ R _0 $ denotes the radius of the uniform filament.   Note that the  
magnetic field surrounding but not permeating the filament is not in the count of 
$ \Phi _{\rm cloud} $.   It should be also noted that $ \Phi _{\rm cloud} $ is twice as
large as $ \Phi _{\rm cl} $ used in Paper I.   Hence, 
the differential mass to flux ratio is expressed as
\begin{eqnarray}
\frac{d\lambda}{d\Phi} & = & \frac{2 \lambda}{\pi \Phi _{\rm cloud}} 
\left[ 1 - \left( \frac{2 \Phi ^\prime}{\Phi _{\rm cloud}} \right) \right] ^{1/2} ,  \label{masstof_tomiska} \\
\Phi ^\prime & \equiv & \int _0 ^x B _z (x ^\prime ) dx  ^\prime .
\end{eqnarray}
Here the symbols, $ B _z (x) $ and $ d\lambda/d\Phi $, denote the magnetic field on the plane of $ z = 0 $
and mass per unit magnetic flux, respectively.   

In order to evaluate the effect of the mass to flux ratio distribution, we modified the cross section of
the initial filament to be
\begin{eqnarray}
\left( \frac{x}{R _0} \right) ^2 + \left( \frac{y}{R _0} \right) ^{2/{\cal N}} & = & 1 ,
\end{eqnarray}
where $ {\cal N} \ge 0 $.   The initial magnetic field is again uniform at $ B _0 $.  
Then the mass to flux ratio is expressed as
\begin{eqnarray}
\frac{d\lambda}{d\Phi} & = & \frac{\Gamma ({\cal N}/2+3/2)}{\sqrt{\pi} \Gamma ({\cal N}/2+1)}
 \frac{\lambda}{\Phi _{\rm cloud}} 
\left[ 1 - \left( \frac{2 \Phi ^\prime}{\Phi _{\rm cloud}} \right) \right] ^{{\cal N}/2} ,  \label{masstof_tomiskaN} 
\end{eqnarray}
where $ \Gamma $ denotes the gamma function.
The mass to flux ratio is uniform in the cloud when $ {\cal N} = 0 $.   On the other hand, 
most of the gas is filled in the magnetic field running at the center in case $ {\cal N} = \infty $.
Thus the parameter, $ {\cal N} $, specifies the mass to flux ratio distribution as well as 
$ w $ in the model shown in \S 3.  See Table \ref{NN} for the ratio of the mass to flux ratio at the center 
to the average. 

\begin{table}
\caption{The ratio of the central mass to flux ratio to the average. \label{NN}}
\begin{center}
\begin{tabular}{rc}
\hline
${\cal N} $ & $ (d\lambda/d\Phi) _c  / (\lambda/ \Phi _{\rm cloud}) $ \\
\hline
0.0 & 1.0000 \\
0.1 & 1.0303 \\
0.2 & 1.0598 \\
1.0 & 1.2732 \\
2.0 & 1.5000 \\
4.0 &  1.8750 \\
10.0 & 2.4610 \\
\hline 
\end{tabular}
\end{center}
\end{table}

\begin{figure}
\epsscale{1.0}
\plottwo{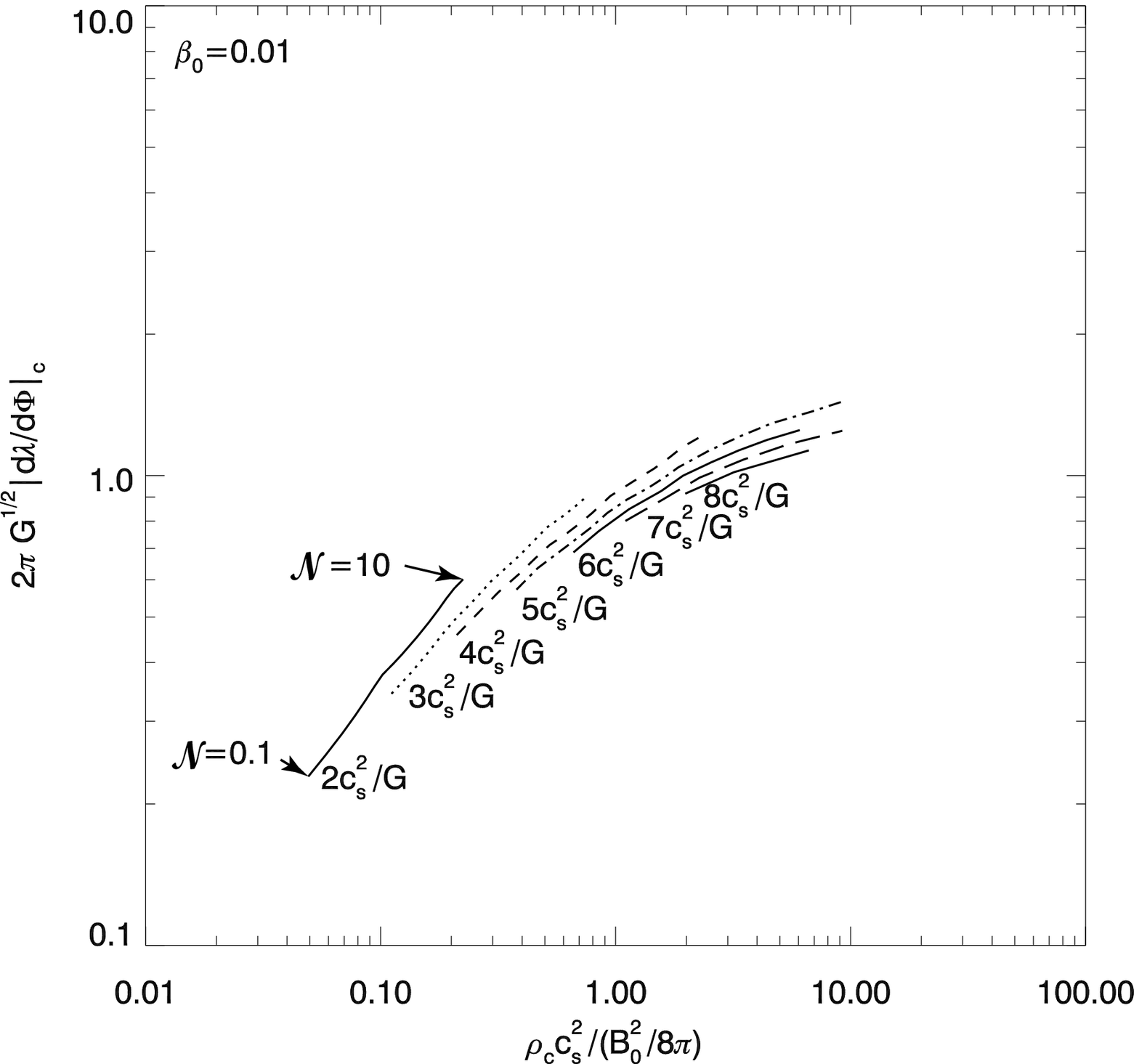}{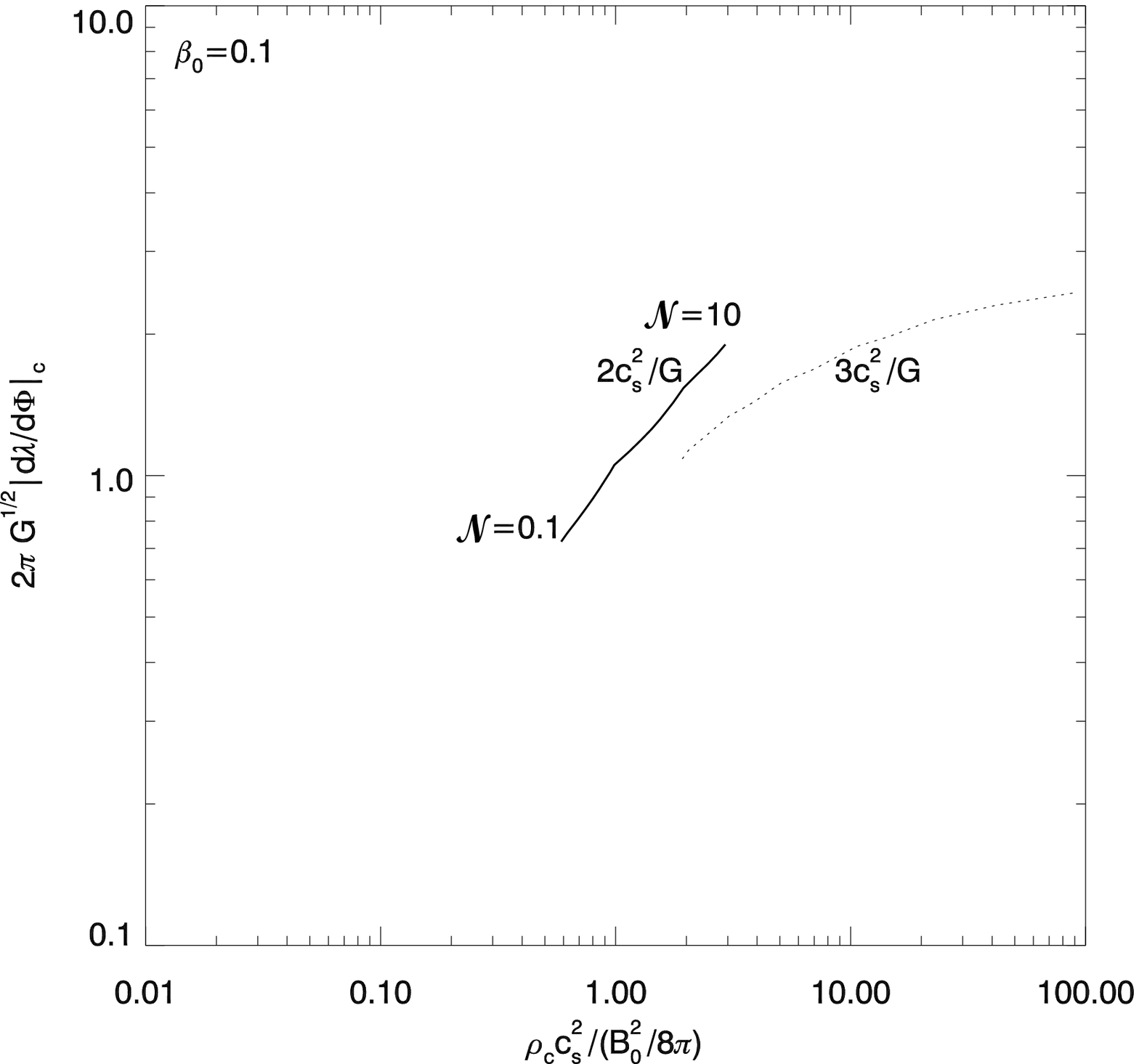}
\caption{The mass to flux ratio at the cloud center  is shown as a function of the density thereof
in 2D models.  The curves denote the loci of constant $ \lambda $ of which values are labelled. 
The central density is denoted in unit of $ B _0 ^2 / 8 \pi c _s ^2 $.
The index, $ {\cal N} $, increases from left to right along the loci.   The left and right panels denote
the models of $ \beta _0 = 0.01 $ and 0.1, respectively.
\label{Sc_ratio2D}}
\end{figure}

Figure \ref{Sc_ratio2D} shows the mass to flux ratio at the cloud center, 
$ 2 \pi \sqrt{G} d\lambda/d\Phi $ as a function of the density at the cloud
center.    Each curve denotes the locus of models having the same $ \lambda $ and $ \beta _0 $,
where $ \beta _0 \equiv \rho _s c _s ^2  (B _0 ^2 / 8 \pi ) ^{-1} $ denotes the initial plasma beta in the ambient gas
surrounding the filamentary cloud.   As in Paper I, the model clouds are assumed to be confined by 
a very tenuous gas of which pressure is $ \rho _s  c _s ^2 $.
The density at the cloud center is measured in unit of $ B _0 ^2 / (8 \pi c _s ^2) $.   
Thus the abscissa denotes the ratio of the gas pressure at the cloud center to the magnetic
pressure in the region very far from the cloud, since the magnetic field is nearly uniform
at the initial value, $ B _0 $, in our 2D model.
The line density is specified in unit of $ c _s ^2 / G $ on the diagram.   

The index, $ {\cal N} $, increases from $ {\cal N} = 0.1 $ (left) to 10 (right) on each locus.    
This means that the filamentary cloud is more centrally condensed and the central mass
to flux ratio is higher when $ {\cal N} $ is larger.   The mass to flux ratio depends only
a little on $ \lambda $ in the models of $ \beta  _0 = 0.01 $.   This is because the gas
is relatively cold and accordingly the gas pressure has only minor contribution to the
cloud support.  Note that the mass to flux ratio is slightly larger for a smaller $\lambda $ 
when $ \rho _c $ is given.   The maximum non-dimensional mass to flux ratio is only slightly 
larger than unity when $ \lambda \ge 5 c _s ^2 / G $ and $ \beta _0 = 0.01 $.

When $\beta _0 = 0.1 $, the gas is relatively warm and hence the pressure has a significant
contribution to the cloud support.  The non-dimensional mass to flux ratio is significantly higher than unity
especially when $  \rho _c >  B _0 ^2 /(8 \pi c _s ^2) $, although it is still less than three.   This high
mass to flux ratio is realized only when the cloud is relatively warm and the line density
is relatively small.

The maximum line density  depends also on $ {\cal N} $, i.e., 
$ \lambda _{\max} = \lambda _{\max}  (\Phi _{\rm cl}, c _s, {\cal N}) $.  Equation (46) of Paper is that 
 for $ {\cal N} = 1 $.  When $ \Phi _{\rm cl} $ and $ c _s $ are given, the maximum line density 
is lower for a larger $ {\cal N} $.  Increase in $ {\cal N} $ mimics  the quasi-static evolution of a molecular
cloud by the ambipolar diffusion.   As $ {\cal N}  $ approaches to zero, the maximum line density tends
to the value given by Equation (\ref{lined}).  

\section{Stability Against Fragmentation}

We consider the stability of the isolated filamentary cloud against fragmentation, i.e., 
a sinusoidal  perturbation in the $ z $-direction.  
The gas is assumed to be still confined in the plane of $ z = 0 $ and 
we use the method of images to obtain the changes in the gravity
and density consistently.   The image density is assumed to have the form,
\begin{eqnarray}
\varrho (x,y,z) & = & \varrho _0 (x,z) + \varrho _1 (x,y,z) , \\
\varrho _0 (x,z) & = & 2 \lambda \delta (x) \delta (z+a) , \\
\varrho _1 (x,y,z) & = & \int _0 ^\infty  \delta \lambda (b) \cos k y \delta (x) \delta (z+b) db ,
\end{eqnarray} 
for calculating the change in the gravity in the upper half space of $ z > 0 $.
The change in the gravity is evaluated as
\begin{eqnarray}
\delta g _z (x,y,z=+\varepsilon) & = & - G \int _{-\infty} ^{+\infty}
\frac{ (z - z ^\prime) \varrho _1 (x^\prime,y^\prime,z^\prime) }
{\left[ (x - x ^\prime) ^2 + (y - y^\prime) ^2 + (z ^\prime) ^2 \right]^{3/2}} 
dx^\prime dy^\prime dz ^\prime \\
& = &  - G b \cos k y \int _{-\infty} ^{+\infty} 
\int _0 ^\infty \frac{ \delta \lambda (b) \cos k s}
{\left( x ^2 + s ^2 + b ^2 \right) ^{3/2}} ds db \\
& = & - 2 G k b \cos k y \int _0 ^\infty \frac{\delta \lambda (b) K _1 \left(k\sqrt{x^2+b^2} \right)}
{\sqrt{x ^2 + b ^2}} db , \label{dgz}
\end{eqnarray}
where $ K _1 $ denotes the modified Bessel function of the 1st order.  Here
we used the mathematical formula \citep{abramowicz65},
\begin{eqnarray}
K _\nu (z) & = & \Gamma \left(\nu + \displaystyle \frac{1}{2} \right)  \frac{\left( 2 z \right) ^\nu}
{\sqrt{\pi}} \int _0 ^\infty \frac{\cos t}{\left( t ^2 + z ^2 \right) ^{\nu+1/2} } dt .
\end{eqnarray}
Similarly we obtain 
\begin{eqnarray}
\delta g _x & = & - 2 G k \cos ky \int _0 ^\infty \frac{x \delta \lambda(b) K _1 \left( k \sqrt{x ^2 + b ^2} \right)}
{\sqrt{x ^2 + b^2}} db ,  \label{dgx} \\
\delta g _y & = &  2 G k \sin ky \int _0 ^\infty \delta \lambda (b) K _0 \left( k \sqrt{x ^2 + b^2} \right) db .
\label{dgy}
\end{eqnarray}
The change in the surface density is expressed as
\begin{eqnarray}
\delta \Sigma (x,y,z=0) & = &
\frac{k b \cos k y}{\pi} \int _0 ^\infty \frac{\delta \lambda (b) K _1 \left(k\sqrt{x^2+b^2} \right)}
{\sqrt{x ^2 + b ^2}} db . \label{dsigma}
\end{eqnarray}

The magnetic field is assumed to be aligned with the gravity also in the perturbed state.
This assumption is reasonable since the gas density is quite low and hence the inertia is 
negligibly small outside the cloud ($ z \ne 0 $).   The magnetic field is still current free and
does not extract any momentum from the cloud.   Then the change in the magnetic field
are expressed as
\begin{eqnarray}
\delta B _x & = &  - \frac{\alpha}{\sqrt{G}} \delta g _x , \\
\delta B _y & = &  - \frac{\alpha}{\sqrt{G}} \delta g _y , \\
\delta B _z & = &  - \frac{\alpha}{\sqrt{G}} \delta g _z ,
\end{eqnarray}
in $ z > 0 $.   Both the gravity and magnetic field change with the time in proportion to 
$ \exp ( - i \omega t ) $, where $ \omega $ denotes the angular frequency of the perturbation.

We use the displacement vector,
\begin{eqnarray}
\mbox{\boldmath$\xi$} & = & \left( \begin{array}{c} \xi _x \cos k y \\ \xi _y \sin k y \end{array} \right) 
\exp ( - i \omega t ) ,
\end{eqnarray}
for our stability analysis.  The $ y $-dependence of the displacement vector is chosen so that
the the equation of mass conservation be the ordinary differential equation with respect to
$ x $,
\begin{eqnarray}
\delta \Sigma + \frac{d}{d x} \left( \xi _x \Sigma _0 \right) + 
k \xi _y \Sigma _0  & = & 0 , \label{per1}
\end{eqnarray}
where $ \Sigma _0 $ denotes the surface density in the equilibrium.
Also the equation of motion reduces to the ordinary differential equation and algebraic equation,
\begin{eqnarray}
- \omega ^2 \Sigma _0 \xi _x +  c _s ^2 \frac{d}{d x} \delta \Sigma 
- \left(1 - \alpha ^2 \right) \left( \Sigma _0 \delta g _x +  g _{x} \delta \Sigma \right) & = & 0 , 
\label{per2} \\
- \omega ^2 \Sigma _0 \xi _y -   c _s ^2 k  \delta \Sigma 
+ \left(1 - \alpha ^2 \right) \Sigma _0 \delta g _y  & = & 0 . \label{per3}
\end{eqnarray}

It is difficult to solve Equations (\ref{per1}) through (\ref{per3}) simultaneously since 
they are associated with Equation (\ref{dgz}), (\ref{dgx}) and  (\ref{dgy}).
Instead of solving the equations we use the variational principle to evaluate $ \omega ^2 $.
First we rewrite Equations (\ref{per2}) and (\ref{per3}) into 
\begin{eqnarray}
- \omega ^2 \xi _x + c _s ^2 \frac{d}{dx} \left( \frac{\delta \Sigma}{\Sigma _0}
\right) - \left(1 - \alpha ^2 \right) \delta g _x & = & 0 , \label{per4} \\
-  \omega ^2 \xi _y + c _s ^2 k \frac{\delta \Sigma}{\Sigma _0} + 
(1 - \alpha ^2 ) \delta g _y & = & 0 ,
\label{per5}
\end{eqnarray}
by using the condition for magnetohydrostatic equilibrium,
\begin{eqnarray}
c _s ^2 \frac{d}{d x} \Sigma _0 - \left(1 - \alpha ^2 \right) \Sigma _0 g _x & = & 0 .
\end{eqnarray}
Note that the displacement is irrotational,
\begin{eqnarray}
\frac{d \xi _y}{d x} + k \xi _x & = & 0 , \label{irrotational}
\end{eqnarray}
since angular momentum extraction by magnetic field is not taken into account.
We obtain
\begin{eqnarray}
- \omega ^2 \Sigma _0 \left( \xi _x ^2 + \xi _y ^2 \right) + c _s ^2 \frac{\delta \Sigma ^2}{\Sigma _0} 
- \left(1 - \alpha ^2 \right) \Sigma _0 \left( \xi _x \delta g _x + \xi _y \delta g _y \right)
+ \frac{d}{d x} \left( c _s ^2 \xi _x \delta \Sigma \right) & = & 0 , \label{var1}
\end{eqnarray}
by taking the sum of the products of $ c _s ^2 \delta \Sigma / \Sigma _0 $ and Equation (\ref{per1}),
$ \xi _x \Sigma _0 $ and Equation (\ref{per4}), $ \xi _y \Sigma _0 $ and Equation (\ref{per5}).
We obtain
\begin{eqnarray}
\omega ^2 
& = & \frac{\displaystyle 
\int _{-\infty} ^{+\infty} \Sigma _0 \left[ c _s ^2 \left(  \frac{\delta \Sigma}{\Sigma _0} \right) ^2  -
\left(1 - \alpha ^2 \right) \left( \xi _x \delta g _x + \xi _y \delta g _y \right)\right] dx  }
{\displaystyle \int _{-\infty} ^{+\infty} \Sigma _0 \left( \xi _x ^2 \, + \, \xi _y ^2 \right) dx}  \\
& = &  c _s ^2 \frac{\displaystyle 
\int _{-\infty} ^{+\infty} \Sigma _0 \left[ \left(  \frac{\delta \Sigma}{\Sigma _0} \right) ^2  -
\frac{1}{G \lambda}  \left( \xi _x \delta g _x  + \xi _y \delta g _y  \right)\right] dx  }
{\displaystyle \int _{-\infty} ^{+\infty} \Sigma _0 \left( \xi _x ^2 \, + \, \xi _y ^2 \right) dx}
\label{var2} 
\end{eqnarray}
by integrating Equation (\ref{var1}) and submitting $ \xi _x = 0 $ at $ x = \pm \infty $.
The right hand side of Equation (\ref{var2}) gives a lower bound for the  
eigenvalue, $ \omega ^2 $,
when an arbitrary perturbation is substituted.  If it is negative for a given perturbation,
the filament is unstable.

We evaluate the right hand side of Equation (\ref{var2}) as a functional of the change in the 
image density, $ \delta \lambda (b) $.  The change in the gravity is evaluated by numerical
integral of (\ref{dgx}), and (\ref{dgy}).   The change in the surface density is evaluated
by the numerical integral of (\ref{dsigma}).  The displacement, $ \xi _x $, is obtained
by solving the ordinary differential equation,
\begin{eqnarray}
\frac{d}{d x} \left( \frac{\delta \Sigma}{\Sigma _0} \right)  + \frac{d}{d x} 
\left[ \frac{1}{\Sigma _0}  \frac{d}{dx} \left( \Sigma _0 \xi _x \right) 
\right] - k ^2 \Sigma _0 \xi _x & = & 0 , \label{xix2}
\end{eqnarray}
which is obtained by combining Equations (\ref{per1}) and (\ref{irrotational}).  The displacement,
$ \xi _y $ is obtained simultaneously when we solve Equation (\ref{xix2}).   
The above mentioned procedure allows us to evaluate the right hand side of Equation (\ref{var2})
as a functional of $ \delta \lambda $.  See Appendix B for more details.

As shown in the previous section, FWHM of
our model cloud (=$a$) is not specified by the condition for magnetohydrostatic equilibrium.
If the filamentary cloud has proper line density and magnetic flux, it can be settled in 
equilibrium at any width.   This means that our model cloud is neutrally stable for 
perturbation having $ k=0 $.    As in the case of longitudinal or helical magnetic field,
we use the non-dimensional wavenumber, $ k a $, in our analysis.

In this paper we consider two types of trial functions for the variational principle.
Type I trial function is expressed as
\begin{eqnarray}
\delta \lambda _1 (b) & = & \delta \left( b - \beta a \right) ,
\end{eqnarray}
where $ \beta $ is chosen to minimize the value of $ - \omega ^2 $.  Type II is
expressed as
\begin{eqnarray}
\delta \lambda _2 (b) & = & \sum _{j=1} ^N \theta _j \delta \left( b - \beta _j a \right) ,
\end{eqnarray} 
where a set of $ \theta _j $ is chosen to minimize the value of $ \omega ^2 $ for
fixed $ \beta _j $.

\begin{figure}
\epsscale{	1.0}
\plottwo{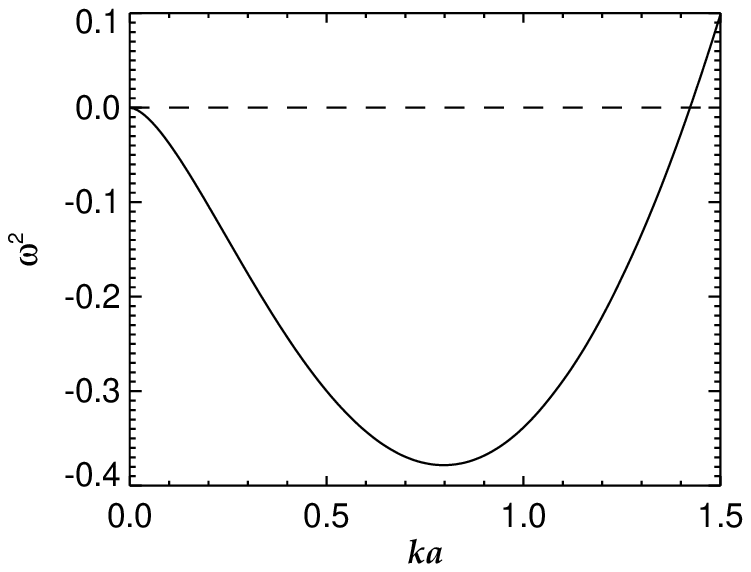}{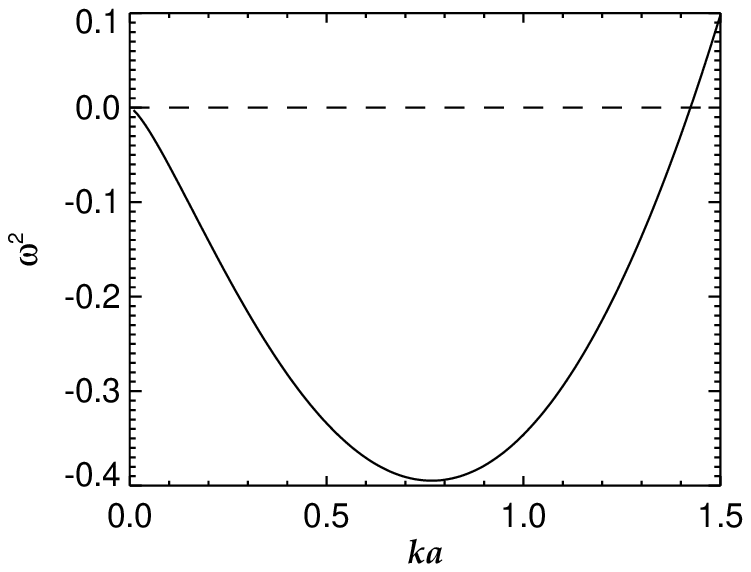}
\caption{The growth rate is shown as a function of the longitudinal wavenumber.
It is derived by applying the type I and type II trial functions to the variational principle,
in left and right panels, respectively.
See text for further details. \label{type1}}
\end{figure}

The left panel of Figure \ref{type1} shows the growth rate obtained by applying type I trial function.
The abscissa is the wavenumber in unit of $ a ^{-1} $ while the ordinate is the growth
rate in unit of $ (c _s/a) ^{2} $.
The right panel of Figure \ref{type1} shows that obtained by applying type II trial function
in which the values of $ \beta _i $ are taken to be 
$ \mbox{\boldmath$\beta$} = (1.0, 1.4, 1.8, 2.2, 2.6) $.    Our model filamentary clouds
are unstable against fragmentation as well as the filamentary clouds threaded by
longitudinal magnetic field.   The most unstable mode has the wave number, 
$ k _{\max} \simeq 0.8 a ^{-1} $, and the growth rate, $ | \omega | \simeq 0.6 c _s a ^{-1} $,
irrespectively of $ \lambda $.   

\section{Discussions and Summary}

As shown in the previous sections, our 1D model based on the thin disk approximation
reproduces the main features of 2D model shown in paper I.   The maximum mass
sustained by magnetic field is roughly proportional to the magnetic flux.   
The critical mass to flux ratio depends on the mass loading, i.e., the distribution of 
$ \Sigma / B $.   When the mass loading is uniform ($ \Sigma / B = $ const.),
it is slightly larger than $ \lambda / B _z > (2 \pi \sqrt{G} ) ^{-1} $.  The same
critical ratio is obtained for disks in equilibrium and slabs stable against fragmentation.  
Remember that \cite{strittmatter66} obtained a similar value from the viral analysis on
the magnetohydrostatic equilibrium.  The critical ratio seems to depend little on the cloud
geometry.   Remember that filamentary clouds with a longitudinal field are unstable
against fragmentation even if the magnetic field is very strong.   The instability is
due to the fact that the mass to flux ratio is infinitely large since the clouds are
highly extended in the direction of magnetic field.

Given the critical mass to flux ratio depends little on the cloud geometry, then it can be
applied to the stability of  periodically arranged filamentary clouds. They are almost
isolated each other in the limit of $ w \rightarrow 0 $.  Thus they are unstable against
fragmentation as well as the isolated filamentary clouds since the mass to flux ratio is
larger than the critical value at the cloud center ($ \Sigma _c > 2 \pi \sqrt{G} B _{z,c} $).
When $ w \rightarrow \infty $, our 1D model reduces to a slab of which stability was 
already investigated by \cite{nakano78}.   The slab is stable as far as 
$ \Sigma  < 2 \pi \sqrt{G} B $.   In short the models shown in Figure \ref{Sc_ratio} are
stable/unstable for small/large value of $ \Sigma _c  $.  We surmise that the models
of $ \Sigma _c  < 2 \pi \sqrt{G} B _{z,c} $ are stable against fragmentation.
Most of the 2D models shown in Figure \ref{Sc_ratio2D} are also likely to be stable
against fragmentation since $ \Sigma _c < 2 \pi \sqrt{G} B _{z,c} $. 

The above argument brings us an interesting result.   Filamentary clouds supported by
longitudinal magnetic fields are unstable against fragmentation while those supported
by perpendicular fields can be stable.   The latter can collapse quasi-statically through
ambipolar diffusion, while the former cannot. 

Our models are still idealistic since turbulence and other dynamical effects are not
taken into account.  Nevertheless, they provide physical insights on the dynamics
of filamentary clouds.   Filamentary clouds with perpendicular magnetic field are
likely to be able to sustain their forms as far as they are subcritical at the cloud
center.   They may fragment after the mass to flux ratio exceeds the critical value
by ambipolar diffusion or by mass accretion along the magnetic field \citep{heitch14}.

This work was supported in part by JSPS KAKENHI Grant Number 24540226.

\appendix

\section{Fourier Series}

We used the mathematical formula,
\begin{eqnarray}
\sum _{n=0} ^\infty a ^n \cos n x & = & \frac{a \cos x - a ^2}{1 - 2 a \cos x + a ^2} ,
\end{eqnarray}
for $ | a | < 1 $ in order to derive Equation (\ref{period1}) from Equations (\ref{psigma}) and (\ref{seriesA}).   
The gravitational potential, $ \phi $, and the $ y $-component of the vector potential, $ A _y $, 
are expressed as
\begin{eqnarray}
\phi & = & \frac{2 \pi G \lambda}{\ell} |z| - \sum _{n=1} ^\infty
\frac{G \lambda}{n} \cos \left( \frac{2 \pi n x}{\ell} \right) \exp \left[ - n
\left( w + \frac{2 \pi}{\ell} |z| \right) \right] , \\
A _y & = & \frac{\Phi x}{\ell} + \sum _{n=1} ^\infty  \frac{\Phi}{\pi n} \sin \frac{2 \pi n x}{\ell} \exp \left[ - n \left( w +
\frac{2 \pi}{\ell} |z| \right) \right] ,
\end{eqnarray}
respectively, for the solution specified by Equations (\ref{period1}) through (\ref{period3}).
They are also expressed as
\begin{eqnarray}
\phi & = & - G \lambda \log \left[ \cosh \left( w + \frac{2 \pi |z|}{\ell} \right)
- \cos \left( \frac{2 \pi x}{\ell} \right) \right] , \\
A _y & = & \frac{\Phi}{\pi} \tan ^{-1} \left[
\frac{\tan \left( \displaystyle \frac{\pi x}{\ell} \right)}
{ \tanh \left( \displaystyle \frac{w}{2} + \displaystyle \frac{\pi |z|}{\ell} \right) }  \right] ,
\end{eqnarray}
respectively, since
\begin{eqnarray}
\sum _{n=1} ^\infty \frac{a ^n}{n} \cos n x & = & - \frac{1}{2} \log
\left( 1 - 2 a \cos x + a ^2 \right) , \\
\sum _{n=1} ^\infty \frac{a ^n}{n} \sin n x & = & \tan ^{-1} \left(
\frac{a \sin x}{1 - a \cos x} \right) ,
\end{eqnarray}
for $ | a | \le 1 $.

\section{Perturbation Equations}

For a given $ \delta \Sigma $ we obtain $ \xi _x $ and $ \xi _y $ by the following procedure.

First we rewrite Equations (\ref{xix2}) and (\ref{irrotational}) as
\begin{eqnarray}
\frac{d \xi _x}{dx} & = & - \frac{d \ln \Sigma _0}{dx} \xi _x -  k \xi _y  
- \frac{\delta \Sigma}{\Sigma _0} ,  \label{per6} \\
\frac{d \xi _y}{dx} & = & - k \xi _x . \label{per7}
\end{eqnarray}  
We consider the case in which $ \delta \Sigma $ 
and $ \xi  _x $ are symmetric and antisymmetric with respect to $ x $, respectively.   
This choice is rational since the unperturbed state is symmetric and we are
interested in the gravitational instability.    An eigenmode perturbation should be either
symmetric or antisymmetric and the fragmentation of a filamentary cloud is symmetric with
respect to $ x $.   Thus we obtain the boundary condition, $ \xi _x = 0 $ at $ x = 0 $.
We assume that the displacement diminishes in the form,
\begin{eqnarray}
\xi _x \propto \exp \left( - k x \right) , \label{asympto}  \\
\xi _y \propto \exp \left(  - k x \right) , \label{asympto2}
\end{eqnarray} 
in the limit of $ x \rightarrow + \infty $.   This choice is based on the fact that 
the relative change in the surface density is proportional to $ \exp ( -  k x) $
in the limit.    Equations (\ref{asympto}) and (\ref{asympto2}) are valid 
also when $ \delta \Sigma / \Sigma _0 $
decreases more steeply with increase in $ x $. 

We integrate Equations (\ref{per6}) and (\ref{per7}) from $ x = 0 $ to $ x _{\rm out}  $ 
with stepsize $ \Delta x = 5 \times 10 ^{-2} $ by the 4th order Runge-Kutta method to
 obtain two sets of solutions.   One has
initial condition $ (\xi _x, d\xi _x/dx) = (0, 0) $.  The other is the solution of Equations
(\ref{per6}) and (\ref{per7})  without the source term, i.e., the homogenous one and has the
initial condition $ (\xi _x, d\xi _x/dx) = (0, 1) $.    We obtain the solution satisfying the
boundary conditions at $ x = 0 $ and $ x _{\rm out} $ as a linear combination of them.
The outer boundary is set at $ x _{\rm out} = \min (15 k ^{-1}, 40) $.
The $ y $-component of the displacement, $ \xi _y$, is derived from Equation (\ref{irrotational}).

Now we can evaluate the right hand side of Equation (\ref{var2}) for a given 
$ \delta \lambda (b) $.   Consequently the growth rate is evaluated to be a function of
$ b $ in type I trial function.  We searched for $ b $ which maximizes the growth rate
in the interval of $ 0.55 a \le b \le  2.5 a $ with the interval $ \Delta b = 0.05 a $.
Both the denominator and delimiter of the right hand side of Equation (\ref{var2}) are
the quadratic expressions of  $ \theta _1, \dots, \theta _5 $, when type II trial function is
used.  We used LAPACK, a mathematical library for linear algebra installed in IDL, the interactive
data language, for minimizing $ - \omega ^2 $.  

We used the public Fortran subroutines downloaded from Prof. Jian-ming Jin's web page
(\verb+jin.ece.illinois.edu+) to obtain the numerical values of  the modified Bessel functions.

\clearpage



\clearpage









\end{document}